\documentclass[conference]{IEEEtran}
\IEEEoverridecommandlockouts

\usepackage{cite}
\usepackage{amsmath,amssymb,amsfonts}
\usepackage{algorithmic}
\usepackage{graphicx}
\usepackage{textcomp}
\usepackage{xcolor}
\usepackage{url}
\usepackage{siunitx}
\usepackage{booktabs, siunitx, tabularx, stfloats}
\usepackage{hyperref}
\usepackage{cleveref}
\usepackage{enumitem}
\usepackage{bm}
\usepackage{flushend}

\crefname{section}{Section}{Sections}
\crefname{figure}{Figure}{Figures}
\crefname{table}{Table}{Tables}

\usepackage{glossaries}
\newacronym{asv}{ASV}{Automatic Speaker Verification}
\newacronym{tts}{TTS}{Text to Speech}
\newacronym{cnn}{CNN}{Convolutional Neural Network}
\newacronym{roc}{ROC}{Receiver Operating Characteristic}
\newacronym{moe}{MoE}{Mixture of Experts}
\newacronym{eer}{EER}{Equal Error Rate}
\newacronym{auc}{AUC}{Area Under the Curve}
\newacronym{mile}{MILE}{Mixture of Implicitly Localized Experts}
\newacronym{mele}{MELE}{Mixture of Explicitly Localized Experts}
\newacronym{msa}{MSA}{Multi-Head Self-Attention}
\newacronym{mlp}{MLP}{Multi-Layer Perceptron}
\newacronym{nlp}{NLP}{Natural Language Processing}

\def\BibTeX{{\rm B\kern-.05em{\sc i\kern-.025em b}\kern-.08em
    T\kern-.1667em\lower.7ex\hbox{E}\kern-.125emX}}
    
\begin{document}

\title{Attention-based Mixture of Experts\\
for Robust Speech Deepfake Detection\\
\thanks{This work was supported by the FOSTERER project, funded by the Italian Ministry of Education, University, and Research within the PRIN 2022 program.
This work was partially supported by the European Union - Next Generation EU under the Italian National Recovery and Resilience Plan (NRRP), Mission 4, Component 2, Investment 1.3, CUP D43C22003080001, partnership on “Telecommunications of the Future” (PE00000001 - program “RESTART”).
This work was partially supported by the European Union - Next Generation EU under the Italian National Recovery and Resilience Plan (NRRP), Mission 4, Component 2, Investment 1.3, CUP D43C22003050001, partnership on ``SEcurity and RIghts in the CyberSpace’’ (PE00000014 - program ``FF4ALL-SERICS’’).}
}

\author{\IEEEauthorblockN{
Viola Negroni,
Davide Salvi,
Alessandro Ilic Mezza,
Paolo Bestagini,
Stefano Tubaro
}\vspace{0.35em}
\IEEEauthorblockA{Dipartimento di Elettronica, Informazione e Bioingegneria (DEIB), Politecnico di Milano, 20133 Milan, Italy\\
\{viola.negroni, davide.salvi, alessandroilic.mezza, paolo.bestagini, stefano.tubaro\}@polimi.it
}
}

\maketitle

\begin{abstract}
AI-generated speech is becoming increasingly used in everyday life, powering virtual assistants, accessibility tools, and other applications. However, it is also being exploited for malicious purposes such as impersonation, misinformation, and biometric spoofing.
As speech deepfakes become nearly indistinguishable from real human speech, the need for robust detection methods and effective countermeasures has become critically urgent.
In this paper, we present the ISPL’s submission to the SAFE challenge at IH\&MMSec 2025, where our system ranked first across all tasks.
Our solution introduces a novel approach to audio deepfake detection based on a Mixture of Experts architecture.
The proposed system leverages multiple state-of-the-art detectors, combining their outputs through an attention-based gating network that dynamically weights each expert based on the input speech signal. In this design, each expert develops a specialized understanding of the shared training data by learning to capture different complementary aspects of the same input through inductive biases.
Experimental results indicate that our method outperforms existing approaches across multiple datasets.
We further evaluate and analyze the performance of our system in the SAFE challenge.

\end{abstract}


\section{Introduction}

We live in an era of unprecedent fast-paced technological advancement, where AI has quickly become integrated into our daily lives and plays a vital role across numerous fields. 
However, this progress does not come without cost. 
As society increasingly relies on AI for routine tasks, its potential for malicious use also grows, leading to potentially severe consequences.
This issue is particularly evident in the domain of media technology. 
With the widespread consumption of short-form, easily digestible content, individuals are more susceptible to mistaking synthetic media for genuine information. 
AI-generated synthetic media, commonly referred to as \textit{deepfakes}, are now frequently used in fraud, reputational damage, and the spread of misinformation \cite{amerini2025deepfake}.

Speech deepfakes, in particular, represent a significant threat, as their misuse for voice phishing, impersonation, extortion, and even biometric spoofing, is raising serious concerns for security and trust in digital communication \cite{li2025survey}.

The forensic community is actively engaged in developing analytical tools and defensive strategies to counter this threat \cite{cuccovillo2024audio, coletta2025anomaly, combei2025unmasking, ge2025post}, and periodic challenges are organized to foster research and promote the deployment of detection systems in simulated real-world scenarios
\cite{yi2023add, wang24_asvspoof, trapeznikov2025safe}.
Nevertheless, most solutions still fall short when required to generalize to unseen deepfake types and data domains.
Also, with the growing complexity of synthetic speech generation, current speech deepfake datasets now differ in key aspects such as generation methods, languages, and post-processing pipelines, effectively making each corpus a distinct domain.
This diversity underscores the urgent need for robust and reliable detection tools capable of generalizing across varied data distributions.

In this work, we present our submission to the Synthetic Audio Forensics Evaluation (SAFE) challenge at IH\&MMSec 2025~\cite{trapeznikov2025safe}. 
The proposed system is based on a \gls{moe} architecture and extends our previous work~\cite{negroni2025leveraging}, aiming to offer a practical contribution to the speech deepfake detection field.

\glspl{moe} can be intended as a dynamic ensembling of multiple specialized models governed by a gating mechanism.
Although dating back to the 1990s~\cite{jacobs1991adaptive}, and being well established in the NLP field~\cite{liu2024deepseek}, their application to deepfake detection remains limited, with only a few studies specifically targeting speech deepfakes~\cite{negroni2025leveraging, wang2025mixture}.
Our \gls{moe}-based approach is designed to improve cross-domain robustness, and introduces two main innovations with respect to previous work: a novel transformer-based gating network, and a revised theoretical framework that better handles unseen domain variability.
The system stochastically partitions the data domain by leveraging the architectural biases of all its distinct deepfake detectors. 
This process is guided by the gating network, which applies attention across the internal representations of each individual expert of the input data.

We validate our method through comparisons with previous work across several state-of-the-art speech deepfake datasets, independently assessing the positive contribution of both the attention-based gating mechanism and the framework shift in a controlled scenario.
We then detail the results obtained in the SAFE challenge, underscoring the validity of our approach in settings that closely reflects real-world deployment scenarios.


\section{Proposed Method}
\label{sec:method}

\begin{figure}
    \centering
    \includegraphics[width=0.9\columnwidth]{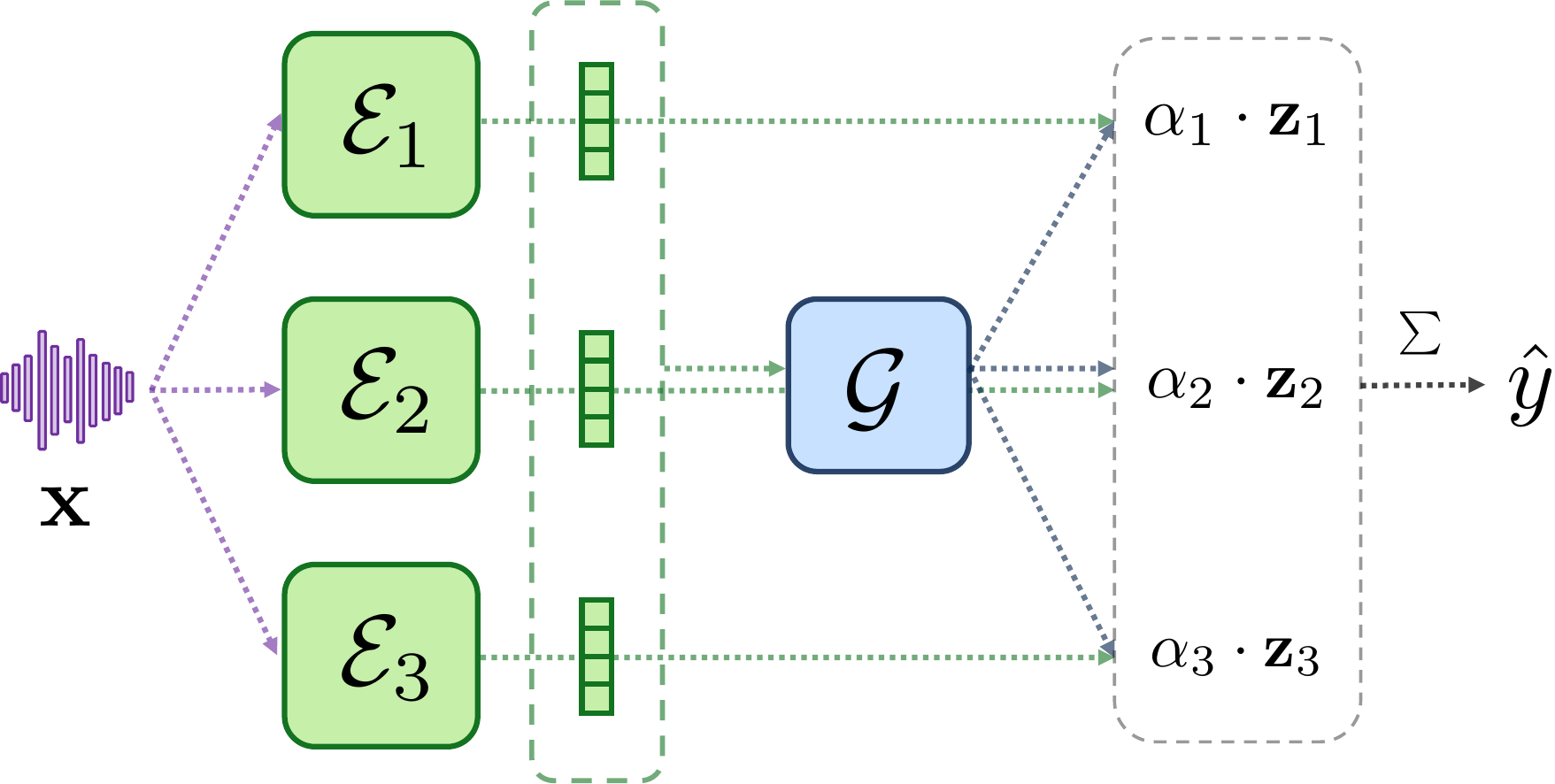}
    \caption{Illustrative example of the proposed \gls{moe} model with $N = 3$ experts. Each expert $\mathcal{E}_i$ processes the input independently, while the gating network $\mathcal{G}$ assigns weights to their outputs based on their internal representations.}
    \label{figure:moe}
\end{figure}

\subsection{Problem Formulation}
\label{subsec:problem}
The speech deepfake detection problem can be formally defined as follows.
Let us consider a discrete-time speech signal $\mathbf{x}$ associated with a class $y \in \{0, 1\}$, where $y = 0$ denotes that the signal is authentic and $y = 1$ indicates that it has been synthetically generated.
The goal of this task is to develop a detector $\mathcal{D}$ that estimates the class of the signal $\mathbf{x}$ as in $\hat{y} = \mathcal{D}(\mathbf{x})$, where $\hat{y} \in [0,1]$ represents the predicted likelihood of $\mathbf{x}$ being fake.

\subsection{Proposed System}
\label{subsec:system}

In this work, we design the detector $\mathcal{D}$ as a \gls{moe} system tailored to the speech deepfake detection task.
Our \gls{moe} framework consists of $N$ expert detectors and a gating network that dynamically weights their outputs to produce a prediction for the given speech input.
We denote the $i$-th expert as $\mathcal{E}_i$, where $i \in \{1, \ldots, N\}$ and $N$ is the total number of experts, each associated with a different architectural inductive bias.
The gating network is denoted as $\mathcal{G}$.
\Cref{figure:moe} shows a simplified illustration of the proposed system.

This system builds on our previous approach presented in~\cite{negroni2025leveraging}, where we introduced a \gls{moe} composed of models with identical architectures, each trained on a distinct speech deepfake domain, with gating governed by a simple concatenation-based mechanism.
Here, we present two key advancements: (1) a novel transformer-based gating architecture, and (2) an improved design involving experts with different architectures being trained on a pooled dataset, to enhance cross-domain robustness.

\vspace{0.2em}
\noindent \textbf{(1) Gating Network Design.}
In the proposed model, the gating network $\mathcal{G}$ is implemented as a transformer encoder with $M$ layers.
This attention-based design enables the gating network to model inter-expert interactions effectively and leverage the representational capacity of transformers.

Each expert $\mathcal{E}_i$ produces an embedding $\mathbf{e}_i$ from the input $\mathbf{x}$, extracted from its final fully connected layer.
These embeddings are linearly projected to a fixed lower-dimensional space $\mathbb{R}^D$ and stacked into a sequence:
\begin{equation}
\mathbf{E}^{(0)} = \left[\mathbf{e}_1, \cdots, \mathbf{e}_N\right]^T \in \mathbb{R}^{N \times D}.
\label{eq:input_gating}
\end{equation}
This serves as input to the transformer, which treats each expert’s internal representation as a separate token and operates accordingly.

Each layer of the transformer encoder, namely $m \in \{1, \ldots, M\}$, contains a \gls{msa} block with $H$ heads, denoted as $\text{MSA}_H(\cdot)$, followed by a feedforward \gls{mlp} block with hidden size $F$, denoted as $\text{MLP}_F(\cdot)$. 
Each layer operates on tensors in $\mathbb{R}^{N \times D}$. 
Layer normalization $\text{LN}(\cdot)$ and residual connections are applied both before and after each block, following the standard transformer configuration~\cite{vaswani2017attention}.

The forward pass through the $m$-th transformer layer is defined as
\begin{align}
\mathbf{A}^{(m)} &= \mathbf{E}^{(m-1)} + \text{MSA}_H\big(\text{LN}(\mathbf{E}^{(m-1)})\big) \\
\mathbf{E}^{(m)} &= \mathbf{A}^{(m)} + \text{MLP}_F\big(\text{LN}(\mathbf{A}^{(m)})\big),
\end{align}
with the final output of the transformer given by 
\begin{equation}
\mathbf{E}^{(M)} \in \mathbb{R}^{N \times D}.
\end{equation}
The output $\mathbf{E}^{(M)}$ is linearly projected to $\mathbb{R}^{N \times 1}$ and then passed through a softmax operation to produce the gating weights $\bm{\alpha}$.

In summary, $G$ yields the experts' weights $\bm{\alpha}$ by attending to each and every input embedding $\mathbf{e}_i$ in $\mathbf{E}^{(0)}$:
\begin{equation}
\bm{\alpha} = \mathcal{G}(\mathbf{E}^{(0)}), \quad \bm{\alpha} \in \mathbb{R}^N.
\end{equation}
Each element $\alpha_i$ in $\bm{\alpha}$ represents the weight assigned to the output of the $i$-th expert.
The final \gls{moe} output $\hat{y}$ is obtained by applying a softmax function to the weighted logits, as in
\begin{equation}
 \label{eq:moe}
 \hat{y} = \operatorname{Softmax}\!\left(\sum_{i=1}^N \alpha_i \cdot \mathbf{z}_i\right),
\end{equation}
where $\mathbf{z}_i$ are the logits of the $i$-th expert for a given input $\mathbf{x}$, as in
\begin{equation}
 \mathbf{z}_i = \mathcal{E}_i(\mathbf{x}).
\end{equation}

\vspace{0.2em}
\noindent \textbf{(2) Domain Partitioning.}
Differently from our previous work~\cite{negroni2025leveraging}, where experts shared the same architecture and were trained on disjoint data domains, the current framework assigns each expert a unique architecture and trains all experts on the same pooled training dataset.
This design encourages each expert to develop a specialized understanding of the data, not by training on separate subsets, but by leveraging its unique architectural biases.
The rationale behind this choice is that specialization should arise from the model's structure rather than from partitioned training domains.

Our design choice aligns with the theoretical distinction between \gls{mile}, as in our case, and \gls{mele}, which better suits the previous framework~\cite{masoudnia2014mixture}.
A \gls{mele} framework performs an explicit partitioning of the problem space before training begins.
Each expert is assigned to a predefined subspace, meaning the decomposition of the problem is determined \textit{a priori}, rather than learned jointly with the experts.
In contrast, a \gls{mile} framework, like the one we propose, induces specialization \textit{during training} and partitions the problem space stochastically into multiple subspaces, with each expert specializing in a different one.
The gating network manages this process and is trained together with the experts in a competitive process.

\section{Experimental Setup}
\label{sec:setup}

This section outlines our experimental setup. 
First, we describe the considered datasets, model architectures, and training strategies. 
Next, we introduce the baseline systems, and finally, we present the challenge-specific setup.

\subsection{Datasets}
\label{subsec:datasets}

We consider seven speech deepfake datasets to rigorously assess both in-domain performance and out-of-domain generalization.
All audio samples are resampled to $\SI{16}{\kilo\hertz}$ for consistency across datasets.

\vspace{0.2em}
\noindent \textbf{ASVspoof 2019~\cite{wang2020asvspoof}}.
Released for the third edition of the homonymous challenge, this dataset includes real speech from the VCTK corpus~\cite{veaux2016superseded} and synthetic speech fabricated by \num{19} different synthetic speech generators. 
We use the Logical Access (LA) subset in our experiments.

\vspace{0.2em}
\noindent \textbf{FakeOrReal~\cite{reimao2019dataset}}.
This dataset consists of both synthetic and real speech samples.
The synthetic (fake) samples are exclusively generated using \gls{tts} systems, comprising a total of \num{7} different open-source and commercial solutions.
The real speech samples, instead, are sourced from publicly available platforms, such as TED Talks and YouTube videos.

\vspace{0.2em}
\noindent \textbf{In-the-Wild~\cite{muller22_interspeech}}. 
This dataset is designed to evaluate speech deepfake detectors in realistic scenarios by providing in-the-wild data. 
It contains an equal balance of fake and real speech samples, featuring \num{54} celebrities and politicians.
The fake clips were generated by segmenting publicly accessible video and audio content, while the real clips were sourced from authentic material featuring the same speakers.

\vspace{0.2em}
\noindent\textbf{MLAAD~\cite{muller2024mlaad}}.
A large-scale multi-lingual dataset of speech deepfakes. 
MLAAD is periodically updated with new generators and languages.
We use version \num{5} of the corpus and restrict our experiments to the English partition. 
This split includes synthetic speech from 36 distinct generators, along with real English speech from M-AILABS~\cite{mailabs_dataset}.

\vspace{0.2em}
\noindent \textbf{Purdue speech dataset~\cite{bhagtani2024recent}}.
This corpus contains synthetic tracks that have been generated using \num{5} advanced diffusion-based voice cloning systems.
The real speech portion includes recordings from LJSpeech~\cite{LJSpeech} and \num{10} speakers from LibriSpeech~\cite{panayotov2015librispeech}. 

\vspace{0.2em}
\noindent \textbf{ASVspoof 2021~\cite{liu2023asvspoof}}.
Released as part of the fourth ASVspoof challenge, 
this dataset did not provide participants with matched training data, making it a purely evaluation-focused dataset.
We utilize the \textit{deepfake} (DF) partition of this dataset, excluding any samples overlapping with ASVspoof 2019.
This includes two additional domains sourced from the 2018 and 2020 Voice Conversion Challenge (VCC) datasets~\cite{lorenzotrueba18_odyssey, yi20_vccbc}, which were not used in earlier ASVspoof challenges.

\vspace{0.2em}
\noindent \textbf{ASVspoof 5~\cite{wang2025asvspoof}}.
Released for the fifth edition of the challenge held in 2024, this dataset was created through a crowdsourcing process, collecting data 
across a wide variety of acoustic environments. 
It features spoofing attacks generated by \num{32} different algorithms, a combination of both legacy and state-of-the-art \gls{tts} synthesis and voice conversion models. 
Notably, it also incorporates adversarial attacks. 
We use only the evaluation set for our experiments.

\vspace{0.2em}
We use \num{4} datasets, i.e., ASVspoof 2019, FakeOrReal, In-the-Wild, and MLAAD, for both training and in-domain testing. 
We follow official training/development/test splits when available; otherwise, we manually split the data to ensure balanced class distributions and coverage of different spoofing methods.

To assess generalization to unseen domains, we evaluate our models on three test-only datasets, i.e., Purdue, ASVspoof 2021, and ASVspoof 5.
For ASVspoof 2021 and ASVspoof 5, we restrict our evaluation to the clean subsets, excluding any audio that has been post-processed with codecs or compression.
This choice allows us to first validate system performance under controlled conditions and ensures a fair comparison with prior work before testing in more challenging, real-world settings such as those simulated in the competition.

\subsection{Models and Training Strategy}
\label{subsec:models_training}

As mentioned in \cref{subsec:system}, rather than assigning each expert to a distinct data subset, we encourage them to focus on complementary aspects of the same data distribution.
The goal of this design is to promote specialization among experts by leveraging architectural and feature-level diversity.
This diversity enables the overall system to capture richer representations, improving both detection performance and robustness to unseen domains.

The selected expert models include three state-of-the-art deepfake detectors:
a LCNN~\cite{wu2018light} and two ResNet18 models~\cite{he2016deep}, all trained using time-frequency representations.
Specifically, one ResNet18 and the LCNN use mel-frequency spectrograms as input, while the second ResNet18 uses linear-frequency spectrograms.
The combination results in three distinct architecture-feature configurations.
This allocation was chosen based on preliminary experiments assessing various architecture-feature pairings.
While those results are not reported here due to space constraints, the chosen configurations offered the best compromise between performance and diversity.
We favored these networks over other established detectors 
due to their favorable balance between model size and training efficiency.
In particular, RawNet2~\cite{tak2021rawnet}, a lightweight model that directly processes raw audio waveforms, 
was also considered; however, its inclusion consistently reduced overall system performance in early tests, which led us to exclude it from the final configuration.

The gating network $\mathcal{G}$ is implemented as a transformer encoder with $M = 2$ layers.
The input to the transformer is a sequence of projected expert embeddings in $\mathbb{R}^{N \times D}$, where $D = 32$ denotes the shared embedding dimension after linear projection.
Each transformer layer uses $H = 4$ attention heads in the \gls{msa} blocks, and an \gls{mlp} block 
of size $F = 512$.

The system processes fixed-length inputs of \SI{4}{\second} of audio, corresponding to \num{64000} samples. 
Inputs shorter than this duration are padded by repeating the signal, while longer utterances are truncated to match the required input length.

Training is conducted in two consecutive stages.
First, we pre-train the individual experts and then we perform a joint training of the full \gls{moe} model, which includes both the experts and the gating network.
For both stages, we use the pooled training and development splits from the datasets introduced in \Cref{subsec:datasets}.
Each expert is pre-trained independently for up to \num{100} epochs, using the Cross-Entropy loss function with \num{20}\% label smoothing. 
We apply early stopping with a patience of \num{20} epochs, monitoring the validation loss. 
Training uses a batch size of \num{256} and the AdamW optimizer with an initial learning rate of \num{e-4}, scheduled via cosine annealing.
Each batch is class-balanced, containing equal numbers of real and fake samples.
After pre-training, the experts (initialized with their trained weights) and the gating network (randomly initialized) are trained jointly. 
This training stage uses the same optimizer and learning rate schedule, but with a reduced batch size of \num{128}, a maximum of \num{50} epochs, and early stopping with a patience of \num{10} epochs.
Training the complete model took less than \num{24} hours on an NVIDIA A40 GPU.

\subsection{Baselines}
\label{subsec:baselines}
To validate the effectiveness of our approach and highlight its advantages over previous work, we compare it against three alternative configurations.

We refer to the attention-based system developed in this work as \textit{\text{MILE}-\text{Att}}. 
\textit{\text{MILE}-\text{Cat}} shares the same framework but uses the fully linear, concatenation-based gating network from~\cite{negroni2025leveraging}.
\textit{\text{MELE}-\text{Cat}} and \textit{\text{MELE}-\text{Att}} both follow the domain partitioning strategy of~\cite{negroni2025leveraging}: 
the first corresponds to the \textit{\gls{moe} Enhanced} model presented in that work, while the second integrates our novel attention-based gating network.
In both \textit{\text{MELE}-\text{Cat}} and \textit{\text{MELE}-\text{Att}}, the experts are four LCNN models fed with mel-frequency spectrograms, each individually pre-trained on a separate training dataset among ASVspoof 2019, FakeOrReal, In-the-Wild, and MLAAD. 
All other training details are consistent with those described in \Cref{subsec:models_training}.

The comparison between \textit{\text{MELE}-\text{Cat}} and \textit{\text{MELE}-\text{Att}} assesses the impact of the new gating function, while the comparison between \textit{\text{MELE}-\text{Att}} and \textit{\text{MILE}-\text{Att}} evaluates the robustness gains from the revised domain partitioning strategy.

\subsection{Challenge System Setup}
\label{subsec:safe_setup}

To achieve the results presented in \Cref{subsec:challenge_res}, we introduced a few enhancements to the model originally designed for the controlled scenario. 
While the system remains the same as described in \Cref{subsec:system}, the technical setup differs slightly from that in \Cref{subsec:models_training}. 
Specifically, we added to the training set real speech samples from LibriSpeech, LJSpeech, and VCTK, beyond those already present in the datasets in \Cref{subsec:datasets}, as well as new data from Mozilla Common Voice~\cite{ardila-etal-2020-common}. 
We also included fake speech samples from the recently released DiffSSD dataset~\cite{bhagtani2025diffssd}. 
During training of both the experts and the gating network, we applied SpecAugment~\cite{park19e_interspeech} and noise augmentations from RawBoost~\cite{tak2022rawboost}.
Since the challenge required the submission of hard (binary) scores, we applied a decision threshold to the model’s output.
The threshold was selected to maximize the balanced accuracy value on the public leaderboard~\cite{trapeznikov2025safe}, and it was empirically set to $0.3$.
\section{Results}
\label{sec:results}
In this section, we present the performance of the proposed system.
In \Cref{subsec:exp_1} and \Cref{subsec:exp_2}, we first conduct two experiments under controlled conditions.
The former experiment is meant to evaluate the superiority of the transformer-based gating network compared to the earlier version in~\cite{negroni2025leveraging}.
The latter validates in terms of robustness the shift in the data domain partitioning strategy among the experts, as described in \Cref{subsec:system}.
Finally, in \Cref{subsec:challenge_res}, we report our system’s performance in the SAFE challenge.

\begin{table*}
\caption{Performance of the proposed attention-based gate \textit{\text{MELE}-\text{Att}} versus the baseline concatenation-based gate \textit{\text{MELE}-\text{Cat}} and their individual experts (LCNN models trained separately on each training dataset). EER (\%), the lower the better.}
\label{tab:exp_1}
\resizebox{\textwidth}{!}{
\begin{tabular}{l|cccc|ccc|ccc}
\hline
\toprule
& \textbf{ASVspoof 2019} & \textbf{FakeOrReal} &\textbf{In-the-Wild} & \textbf{MLAAD} & \textbf{Purdue} & \textbf{ASVspoof 2021} & \textbf{ASVspoof 5} & \textbf{Known} & \textbf{Unknown} & \textbf{Overall} \\ \midrule \midrule
$\mathcal{E}_\text{ASV19}$   & 9.11           & 5.70           & 39.08          & 43.78          & 33.84          & 31.43          & 29.43          & 24.42          & 31.44          & 27.43 \\
$\mathcal{E}_\text{FOR}$       & 27.90          & 6.05           & 20.44          & 46.49          & 11.53          & 34.29          & 35.54          & 25.22          & 27.97          & 26.40 \\
$\mathcal{E}_\text{ITW}$       & 21.65          & 2.03           & \textbf{0.40}  & 35.31          & 13.61          & 43.64          & 37.13          & 14.85          & 30.94          & 21.75 \\
$\mathcal{E}_\text{MLA}$       & 27.97          & 61.84          & 23.91          & 1.56           & 60.77          & 60.91          & 30.35          & 28.82          & 50.67          & 38.18 \\
\textit{\text{MELE}-\text{Cat}}                       & 7.37           & 2.69           & 1.17           & 1.24           & \textbf{3.88}  & 31.82          & \textbf{15.00} & 3.12           & 16.31          & 8.77  \\
\textit{\text{MELE}-\text{Att}}                       & \textbf{7.25}  & \textbf{1.90}  & 1.10           & \textbf{1.02}  & 4.53           & \textbf{29.87} & 15.34          & \textbf{2.82}  & \textbf{15.75} & \textbf{8.36}  \\ \bottomrule 
\end{tabular}}
\end{table*}

\begin{table*}
\caption{
Performance impact of domain partitioning. 
Comparison of the proposed system
\textit{\text{MILE}-\text{Att}} (pooled-data diverse experts),
\textit{\text{MELE}-\text{Att}} (separate-domain LCNNs), and \textit{\text{MILE}-\text{Cat}} (pooled-data diverse experts, baseline gate). EER (\%), the lower the better.
}
\label{tab:exp_2}
\resizebox{\textwidth}{!}{
\begin{tabular}{l|cccc|ccc|ccc}
\hline
\toprule
& \textbf{ASVspoof 2019} & \textbf{FakeOrReal} &\textbf{In-the-Wild} & \textbf{MLAAD} & \textbf{Purdue} & \textbf{ASVspoof 2021} & \textbf{ASVspoof 5} & \textbf{Known} & \textbf{Unknown} & \textbf{Overall} \\ \midrule \midrule
\textit{\text{MILE}-\text{Att}}   & 8.38          & 0.04           & 0.57           & \textbf{0.04} & 6.49          & \textbf{17.27} & 17.69          & \textbf{2.26} & \textbf{14.15} & \textbf{7.35}  \\
\textit{\text{MELE}-\text{Att}}   & \textbf{7.25} & 1.90           & 1.10           & 1.02          & \textbf{4.53} & 29.87          & \textbf{15.34} & 2.82          & 15.75          & 8.36           \\
\textit{\text{MILE}-\text{Cat}}   & 8.86          & \textbf{0.00}  & \textbf{0.37}  & 0.07          & 10.68         & 18.18          & 18.72          & 2.32          & 18.09          & 9.08           \\ \bottomrule 
\end{tabular}}
\end{table*}

\begin{table}
\caption{Performance of the proposed system and of the individual experts on Task 1 (public leaderboard), reported as percentage values for TPR, TNR, and BAC.}
\label{tab:task_1}
\centering
\begin{tabular}{lccc}
\hline
\toprule
           & TPR & TNR & BAC \\ \midrule \midrule
$\mathcal{E}_\text{1}$ (Mel - LCNN)   & \textbf{86.4} & 79.9          & 83.2   \\
$\mathcal{E}_\text{2}$ (Mel - ResNet18)   & 83.6          & 86.6          & 85.1   \\
$\mathcal{E}_\text{3}$ (Linear - ResNet18)   & 83.7          & 91.6          & 87.7   \\ 
\textit{\text{MILE}-\text{Att}}   & 85.9          & \textbf{93.3} & \textbf{89.6}   \\ \bottomrule 
\end{tabular}
\end{table}

\begin{table}
\caption{Performance of the proposed system on Tasks 2 and 3 of the SAFE Challenge (public leaderboard), reported as percentage values for TPR, TNR, and BAC.}
\label{tab:task_2_3}
\centering
\begin{tabular}{lccc}
\hline
\toprule
           & TPR & TNR & BAC \\ \midrule \midrule
Task 2 (Post-processing)     & 76.3   & 88.2   & 82.2   \\
Task 3  (Laundering)   & 41.1   & 87.8   & 64.5   \\ \bottomrule 
\end{tabular}
\end{table}

\subsection{Impact of Gating Network Design}
\label{subsec:exp_1}
This experiment evaluates the first of the two main contributions presented in this work, namely the proposed attention-based gating network. 

Here, we adopt the same framework as in~\cite{negroni2025leveraging}, where identical expert models are trained independently on each training dataset (see \Cref{subsec:baselines}).
In \Cref{tab:exp_1}, \textit{\text{MELE}-\text{Cat}} refers to the best-performing \gls{moe} from~\cite{negroni2025leveraging}, while \textit{\text{MELE}-\text{Att}} denotes the same system enhanced with the proposed attention-based gating function. 
We report EER (\%) for both \glspl{moe}, along with the performance of their four constituent experts. Results are provided for each known and unknown domain, as well as aggregated into \textit{known}, \textit{unknown}, and \textit{overall} metrics. \textit{\text{MELE}-\text{Att}} consistently outperforms both \textit{\text{MELE}-\text{Cat}} and the individual experts across all aggregate metrics, demonstrating the effectiveness of the proposed gating mechanism.
Interestingly, not every expert performs best on its corresponding domain (e.g., $\mathcal{E}_\text{FOR}$), and some, such as $\mathcal{E}_\text{ITW}$ and $\mathcal{E}_\text{MLA}$, show strong performance on their own domain but struggle significantly on others. 
These variations may stem from a range of factors that are not always predictable or easily controlled, and highlight the strength of \glspl{moe} in effectively handling such domain-specific inconsistencies, by dynamically assigning more weight to the most suitable expert(s), regardless of a priori domain assignment.

\subsection{Impact of Domain Partitioning Strategy}
\label{subsec:exp_2}
In this experiment, we assess the second main contribution of this work: the shift from using the same expert model trained on separate domains to using different expert networks trained on pooled data.

\Cref{tab:exp_2} presents the results for the proposed system, \textit{\text{MILE}-\text{Att}}, alongside its counterpart \textit{\text{MELE}-\text{Att}} (from the experiment in \Cref{subsec:exp_1}), which uses the same gating network but adopts the old domain partitioning strategy. 
We also report results for another variant, \textit{\text{MILE}-\text{Cat}}, which adopts the new domain partitioning strategy but retains the old gating network from~\cite{negroni2025leveraging}.
\textit{\text{MILE}-\text{Att}} outperforms all its counterparts across all aggregate metrics: known domains, unknown domains, and overall.
Interestingly enough, while all \gls{moe} variants perform strongly on known domains, those following the new framework achieve a relative reduction in EER of nearly 10 percentage points on the unknown ASVspoof 2021 domain, which proved particularly challenging in the previous experiment (\Cref{tab:exp_1}).
These results underscore the effectiveness of the proposed domain partitioning strategy in enhancing each expert’s inductive biases, leading to improved generalization.

\subsection{Challenge results}
\label{subsec:challenge_res}
Here, we present the results of our \textit{\text{MILE}-\text{Att}} \gls{moe} system in the SAFE challenge~\cite{trapeznikov2025safe}. 
The challenge consisted of three tasks, with no data provided for training or independent evaluation. 
Task 1 involved real and deepfake data without post-processing, Task 2 included synthetic audio post-processed with effects such as codecs and noise, while Task 3 involved laundered synthetic audio. 
The real data were left untouched in every scenario.
All systems were ranked exclusively by means of the balanced accuracy metric.

\Cref{tab:task_1} presents the results for Task~1, where our system secured first place on both the public and private leaderboards. 
Performance metrics include the Balanced Accuracy (BAC), True Positive Rate (TPR) for fake detection, and True Negative Rate (TNR) for real detection. 
Results for the individual experts are also included, as Task 1 served to calibrate the final model deployed for Tasks 2 and 3.
\Cref{tab:task_2_3} reports the results for the remaining tasks. 
For Task 2, we achieved first place on both the public and private leaderboards, while for Task 3, we ranked fourth on the public leaderboard and first on the private leaderboard.	

Our \gls{moe} framework demonstrated strong performance on Task 1, achieving a balanced accuracy of \num{89.6}\%. 
This is especially notable given that the challenge policy neither included the release of any data nor provided any details in this regard.
On the other hand, while the system maintained reasonable robustness on Task 2, there is substantial room for improvement on Task 3.

The performance degradation observed between the tasks highlights some important limitations of the current system.  
Balanced accuracy drops by \num{7.4}\% from Task 1 to Task 2, and by a more pronounced \num{25.1}\% from Task 1 to Task 3. 
This suggests that the data augmentations considered during training, while effective in allowing for a strong generalization on Task 1, might not have been sufficient to guarantee broader robustness under more challenging conditions. 
The limitation is particularly evident in Task 3, where synthetic audio was augmented with car background noise, reverberation, or played over the air, creating much more realistic and complex distortions that the system struggled to handle.

\section{Conclusions}
\label{sec:conclusion}

In this work, we presented ISPL’s submission to the SAFE challenge at IH\&MMSec 2025, where our system ranked first across all three tasks proposed by the organizers.
Building on prior work, we introduced a novel approach to audio deepfake detection based on a \gls{moe} architecture. 
New contributions include an attention-based gating mechanism and a redesigned data domain partitioning strategy, which were thoroughly validated through targeted experiments.
Looking ahead, we aim to explore alternative \gls{moe} configurations and more expressive expert architectures. 
We also plan to further strengthen robustness against post-processing operations.
At the same time, we acknowledge that \gls{moe} frameworks can lead to significant computational costs. 
While our submitted system was the fastest among the top five performers on Task 1, future work will also focus on assessing the scalability of our approach and addressing the practical constraints given by resource consumption in real-world deployments.

\bibliographystyle{IEEEtran}
\bibliography{bstcontrol.bib, bibliography.bib}

\begin{thebibliography}{10}
\providecommand{\url}[1]{#1}
\csname url@samestyle\endcsname
\providecommand{\newblock}{\relax}
\providecommand{\bibinfo}[2]{#2}
\providecommand{\BIBentrySTDinterwordspacing}{\spaceskip=0pt\relax}
\providecommand{\BIBentryALTinterwordstretchfactor}{4}
\providecommand{\BIBentryALTinterwordspacing}{\spaceskip=\fontdimen2\font plus
\BIBentryALTinterwordstretchfactor\fontdimen3\font minus \fontdimen4\font\relax}
\providecommand{\BIBforeignlanguage}[2]{{%
\expandafter\ifx\csname l@#1\endcsname\relax
\typeout{** WARNING: IEEEtran.bst: No hyphenation pattern has been}%
\typeout{** loaded for the language `#1'. Using the pattern for}%
\typeout{** the default language instead.}%
\else
\language=\csname l@#1\endcsname
\fi
#2}}
\providecommand{\BIBdecl}{\relax}
\BIBdecl

\bibitem{amerini2025deepfake}
I.~Amerini, M.~Barni, S.~Battiato, P.~Bestagini, G.~Boato, V.~Bruni, R.~Caldelli, F.~De~Natale, R.~De~Nicola, L.~Guarnera \emph{et~al.}, ``Deepfake media forensics: Status and future challenges,'' \emph{Journal of Imaging}, vol.~11, no.~3, p.~73, 2025.

\bibitem{li2025survey}
M.~Li, Y.~Ahmadiadli, and X.-P. Zhang, ``A survey on speech deepfake detection,'' \emph{ACM Computing Surveys}, 2025.

\bibitem{cuccovillo2024audio}
L.~Cuccovillo, M.~Gerhardt, and P.~Aichroth, ``Audio transformer for synthetic speech detection via multi-formant analysis,'' in \emph{IEEE/CVF Conference on Computer Vision and Pattern Recognition Workshops (CVPRW)}, 2024.

\bibitem{coletta2025anomaly}
E.~Coletta, D.~Salvi, V.~Negroni, D.~U. Leonzio, and P.~Bestagini, ``Anomaly detection and localization for speech deepfakes via feature pyramid matching,'' in \emph{European Signal Processing Conference (EUSIPCO)}, 2025.

\bibitem{combei2025unmasking}
D.~Combei, A.~Stan, D.~Oneata, N.~M{\"u}ller, and H.~Cucu, ``Unmasking real-world audio deepfakes: A data-centric approach,'' in \emph{Interspeech}, 2025.

\bibitem{ge2025post}
W.~Ge, X.~Wang, X.~Liu, and J.~Yamagishi, ``{Post-training for Deepfake Speech Detection},'' \emph{arXiv preprint arXiv:2506.21090}, 2025.

\bibitem{yi2023add}
J.~Yi, J.~Tao, R.~Fu, X.~Yan, C.~Wang, T.~Wang, C.~Zhang, X.~Zhang, Y.~Zhao, Y.~Ren, L.~Xu, J.~Zhou, H.~Gu, Z.~Wen, S.~Liang, Z.~Lian, S.~Nie, and H.~Li, ``Add 2023: the second audio deepfake detection challenge,'' \emph{CEUR Workshop Proceedings}, 2023.

\bibitem{wang24_asvspoof}
X.~Wang, H.~Delgado, H.~Tak, J.~weon Jung, H.~jin Shim, M.~Todisco, I.~Kukanov, X.~Liu, M.~Sahidullah, T.~H. Kinnunen, N.~Evans, K.~A. Lee, and J.~Yamagishi, ``Asvspoof 5: crowdsourced speech data, deepfakes, and adversarial attacks at scale,'' in \emph{The Automatic Speaker Verification Spoofing Countermeasures Workshop}, 2024.

\bibitem{trapeznikov2025safe}
T.~Kirill, P.~Cummer, P.~Pherwani, J.~Aslam, M.~Davinroy, P.~Bautista, L.~Cassani, and M.~Stamm, ``{SAFE: Synthetic Audio Forensics Evaluation Challenge},'' in \emph{Association for Computing Machinery, IH\&MMSEC}, 2025.

\bibitem{negroni2025leveraging}
V.~Negroni, D.~Salvi, A.~I. Mezza, P.~Bestagini, and S.~Tubaro, ``Leveraging mixture of experts for improved speech deepfake detection,'' in \emph{IEEE International Conference on Acoustics, Speech and Signal Processing (ICASSP)}, 2025.

\bibitem{jacobs1991adaptive}
R.~A. Jacobs, M.~I. Jordan, S.~J. Nowlan, and G.~E. Hinton, ``Adaptive mixtures of local experts,'' \emph{Neural computation}, vol.~3, no.~1, pp. 79--87, 1991.

\bibitem{liu2024deepseek}
A.~Liu, B.~Feng, B.~Xue, B.~Wang, B.~Wu, C.~Lu, C.~Zhao, C.~Deng, C.~Zhang, C.~Ruan \emph{et~al.}, ``Deepseek-v3 technical report,'' \emph{arXiv preprint arXiv:2412.19437}, 2024.

\bibitem{wang2025mixture}
Z.~Wang, R.~Fu, Z.~Wen, J.~Tao, X.~Wang, Y.~Xie, X.~Qi, S.~Shi, Y.~Lu, Y.~Liu \emph{et~al.}, ``Mixture of experts fusion for fake audio detection using frozen wav2vec 2.0,'' in \emph{IEEE International Conference on Acoustics, Speech and Signal Processing (ICASSP)}, 2025.

\bibitem{vaswani2017attention}
A.~Vaswani, N.~Shazeer, N.~Parmar, J.~Uszkoreit, L.~Jones, A.~N. Gomez, {\L}.~Kaiser, and I.~Polosukhin, ``Attention is all you need,'' \emph{Advances in neural information processing systems}, 2017.

\bibitem{masoudnia2014mixture}
S.~Masoudnia and R.~Ebrahimpour, ``Mixture of experts: a literature survey,'' \emph{Artificial Intelligence Review}, pp. 275--293, 2014.

\bibitem{wang2020asvspoof}
X.~Wang, J.~Yamagishi, M.~Todisco, H.~Delgado, A.~Nautsch, N.~Evans, M.~Sahidullah, V.~Vestman, T.~Kinnunen, K.~A. Lee \emph{et~al.}, ``{ASVspoof 2019: A large-scale public database of synthesized, converted and replayed speech},'' \emph{Computer Speech \& Language}, p. 101114, 2020.

\bibitem{veaux2016superseded}
C.~Veaux, J.~Yamagishi, K.~MacDonald \emph{et~al.}, ``{Superseded-CSTR VCTK Corpus: English Multi-Speaker Corpus for CSTR Voice Cloning Toolkit},'' \emph{University of Edinburgh. The Centre for Speech Technology Research (CSTR)}, 2016.

\bibitem{reimao2019dataset}
R.~Reimao and V.~Tzerpos, ``{FoR: A dataset for synthetic speech detection},'' in \emph{IEEE International Conference on Speech Technology and Human-Computer Dialogue (SpeD)}, 2019.

\bibitem{muller22_interspeech}
N.~Müller, P.~Czempin, F.~Diekmann, A.~Froghyar, and K.~Böttinger, ``{Does Audio Deepfake Detection Generalize?}'' in \emph{Interspeech}, 2022.

\bibitem{muller2024mlaad}
N.~M. M{\"u}ller, P.~Kawa, W.~H. Choong, E.~Casanova, E.~G{\"o}lge, T.~M{\"u}ller, P.~Syga, P.~Sperl, and K.~B{\"o}ttinger, ``{MLAAD: The multi-language audio anti-spoofing dataset},'' in \emph{IEEE International Joint Conference on Neural Networks (IJCNN)}, 2024.

\bibitem{mailabs_dataset}
{Celeste, Imdat, et al.}, ``M-ailabs speech dataset,'' \url{https://github.com/imdatceleste/m-ailabs-dataset}, accessed: 2025-06-30.

\bibitem{bhagtani2024recent}
K.~Bhagtani, A.~K.~S. Yadav, P.~Bestagini, and E.~J. Delp, ``{Are Recent Deepfake Speech Generators Detectable?}'' in \emph{ACM Workshop on Information Hiding and Multimedia Security}, 2024.

\bibitem{LJSpeech}
K.~Ito and L.~Johnson, ``{The LJ Speech Dataset},'' \url{https://keithito.com/LJ-Speech-Dataset/}, 2017.

\bibitem{panayotov2015librispeech}
V.~Panayotov, G.~Chen, D.~Povey, and S.~Khudanpur, ``{Librispeech: an ASR corpus based on public domain audio books},'' in \emph{IEEE International Conference on Acoustics, Speech and Signal Processing (ICASSP)}, 2015.

\bibitem{liu2023asvspoof}
X.~Liu, X.~Wang, M.~Sahidullah, J.~Patino, H.~Delgado, T.~Kinnunen, M.~Todisco, J.~Yamagishi, N.~Evans, A.~Nautsch \emph{et~al.}, ``{ASVspoof 2021: Towards spoofed and deepfake speech detection in the wild},'' \emph{IEEE ACM Transactions on Audio, Speech, and Language Processing}, pp. 2507--2522, 2023.

\bibitem{lorenzotrueba18_odyssey}
J.~Lorenzo-Trueba, J.~Yamagishi, T.~Toda, D.~Saito, F.~Villavicencio, T.~Kinnunen, and Z.~Ling, ``{The Voice Conversion Challenge 2018: Promoting Development of Parallel and Nonparallel Methods},'' in \emph{The Speaker and Language Recognition Workshop (Odyssey)}, 2018.

\bibitem{yi20_vccbc}
Z.~Yi, W.-C. Huang, X.~Tian, J.~Yamagishi, R.~K. Das, T.~Kinnunen, Z.-H. Ling, and T.~Toda, ``{Voice Conversion Challenge 2020 –- Intra-lingual semi-parallel and cross-lingual voice conversion –-},'' in \emph{Joint Workshop for the Blizzard Challenge and Voice Conversion Challenge}, 2020.

\bibitem{wang2025asvspoof}
X.~Wang, H.~Delgado, H.~Tak, J.-w. Jung, H.-j. Shim, M.~Todisco, I.~Kukanov, X.~Liu, M.~Sahidullah, T.~Kinnunen \emph{et~al.}, ``{ASVspoof 5: Design, collection and validation of resources for spoofing, deepfake, and adversarial attack detection using crowdsourced speech},'' \emph{Computer Speech \& Language}, p. 101825, 2025.

\bibitem{wu2018light}
X.~Wu, R.~He, Z.~Sun, and T.~Tan, ``{A light CNN for deep face representation with noisy labels},'' \emph{{IEEE Transactions on Information Forensics and Security}}, pp. 2884--2896, 2018.

\bibitem{he2016deep}
K.~He, X.~Zhang, S.~Ren, and J.~Sun, ``Deep residual learning for image recognition,'' in \emph{IEEE conference on Computer Vision and Pattern Recognition (CVPR)}, 2016.

\bibitem{tak2021rawnet}
H.~Tak, J.~Patino, M.~Todisco, A.~Nautsch, N.~Evans, and A.~Larcher, ``End-to-end anti-spoofing with rawnet2,'' in \emph{IEEE International Conference on Acoustics, Speech and Signal Processing (ICASSP)}, 2021.

\bibitem{ardila-etal-2020-common}
R.~Ardila, M.~Branson, K.~Davis, M.~Kohler, J.~Meyer, M.~Henretty, R.~Morais, L.~Saunders, F.~Tyers, and G.~Weber, ``Common voice: A massively-multilingual speech corpus,'' in \emph{Proceedings of the Twelfth Language Resources and Evaluation Conference}, 2020.

\bibitem{bhagtani2025diffssd}
K.~Bhagtani, A.~K.~S. Yadav, P.~Bestagini, and E.~J. Delp, ``Diffssd: A diffusion-based dataset for speech forensics,'' in \emph{IEEE International Conference on Acoustics, Speech and Signal Processing (ICASSP)}, 2025.

\bibitem{park19e_interspeech}
D.~S. Park, W.~Chan, Y.~Zhang, C.-C. Chiu, B.~Zoph, E.~D. Cubuk, and Q.~V. Le, ``{SpecAugment: A Simple Data Augmentation Method for Automatic Speech Recognition},'' in \emph{Interspeech}, 2019.

\bibitem{tak2022rawboost}
H.~Tak, M.~Kamble, J.~Patino, M.~Todisco, and N.~Evans, ``{Rawboost: A raw data boosting and augmentation method applied to automatic speaker verification anti-spoofing},'' in \emph{IEEE International Conference on Acoustics, Speech and Signal Processing (ICASSP)}, 2022.

\end{thebibliography}

\end{document}